\documentclass[11pt,a4paper]{article}
\usepackage[hyperref]{acl2020}
\usepackage{times}
\usepackage{latexsym}
\usepackage{graphicx}
\usepackage{booktabs}
\usepackage{url}
\usepackage{CJK}

\newcommand{\ignore}[1]{}

\usepackage{microtype}

\aclfinalcopy % Uncomment this line for the final submission
\title{Cross-Lingual Relevance Transfer for Document Retrieval}

\author{Peng Shi \and Jimmy Lin\vspace{0.1cm}\\
  David R. Cheriton School of Computer Science \\
  University of Waterloo}

\date{}

\begin{document}
\maketitle
\begin{abstract}
Recent work has shown the surprising ability of multi-lingual BERT to serve as a zero-shot cross-lingual transfer model for a number of language processing tasks.
We combine this finding with a similarly-recently proposal on sentence-level relevance modeling for document retrieval to demonstrate the ability of multi-lingual BERT to transfer models of relevance across languages.
Experiments on test collections in five different languages from diverse language families (Chinese, Arabic, French, Hindi, and Bengali) show that models trained with English data improve ranking quality, without any special processing, both for (non-English) mono-lingual retrieval as well as cross-lingual retrieval.
\end{abstract}

\section{Introduction}

Transformer models that have been pretrained on language modeling tasks such as BERT~\citep{devlin2018bert} have led to many advances in diverse language processing tasks ranging from textual inference to sequence labeling.
Interest in these models have also extended to search-related tasks such as retrieval-based question answering~\cite{Yang_etal_NAACL2019demo}, passage ranking~\cite{nogueira2019passage}, and document ranking~\cite{Yang_etal_arXiv2019b,MacAvaney_etal_SIGIR2019,Yilmaz_etal_EMNLP2019}.

Our work builds on~\citet{Yilmaz_etal_EMNLP2019}, who proposed a simple approach to document ranking that aggregates sentence-level evidence (based on BERT) with document-level evidence (based on traditional exact term-matching scores).
Furthermore, they demonstrated that BERT models fine-tuned with passage-level relevance data can transfer across domains:\ surprisingly, fine-tuning on social media data is effective for relevance classification on newswire documents without any additional modifications.

Inspired by the work of~\citet{wu-dredze-2019-beto}, who explored the cross-lingual potential of multi-lingual BERT (henceforth, mBERT for short) as a zero-shot language transfer model, we wondered if the techniques of~\citet{Yilmaz_etal_EMNLP2019} would transfer across languages in addition to transferring across domains.
Supported by experiments in five different non-English languages from diverse language families (Chinese, Arabic, French, Hindi, and Bengali)---we find, perhaps unsurprisingly, the answer is {\it yes}!

The contribution of this work is empirical validation that the cross-domain relevance transfer work of~\citet{Yilmaz_etal_EMNLP2019} also works {\it cross-lingually} without any additional effort, for both mono-lingual retrieval in non-English languages as well as cross-lingual retrieval.
We demonstrate robust increases in document retrieval effectiveness across diverse languages that come ``for free''.

\section{Background and Approach}

Our work adopts the standard formulation of document ranking:\ given a user query $Q$, the system's task is to produce a ranking of documents from a corpus that maximizes some ranking metric---in our case, average precision (AP).
In the context of cross-lingual transfer learning, it is useful to precisely define the {\it source} language (the language of the training data) and the {\it target} language (the language in which inference is being applied).
In our case, the source language is English.
There are two variants of our retrieval task:\ In mono-lingual target language retrieval, both the query and the documents are in another language (for example, Bengali).
In cross-lingual retrieval, the query and the documents are in different languages (for example, English queries, Bengali documents).

Following~\citet{wu-dredze-2019-beto}, we use mBERT, which has been pretrained on concatenated Wikipedia data for 104 languages, as our transfer model.
Starting with mBERT, we fine-tune the model for sentence-level relevance classification as described by~\citet{Yilmaz_etal_EMNLP2019}, which is based on~\citet{nogueira2019passage}.

Starting from pretrained mBERT, we fine-tune the model as follows:\ the input to mBERT comprises [\texttt{[CLS]}, $Q$  \texttt{[SEP]} $S$ \texttt{[SEP]}], which is the concatenation of the query $Q$ and a sentence $S$, with the standard special tokens \texttt{[CLS]} and \texttt{[SEP]}.
The final hidden state of the \texttt{[CLS]} token is passed to a single layer neural network with a softmax, obtaining the probability that sentence $S$ is relevant to the query $Q$. 

Following~\citet{Yilmaz_etal_EMNLP2019}, the model (mBERT in our case) is fine-tuned with data from the TREC Microblog Tracks~\cite{Lin_etal_TREC2014}, since typical IR test collections---which only have relevance annotated at the document level---are too long for feeding into mBERT.
Despite the mismatch in domain between training data and test data (tweets vs.\ newswire documents), the previous work showed that relevance matching models transfer across domains.

For document retrieval (i.e., at inference time), let us first consider the case of mono-lingual retrieval in the target language (i.e., queries in Bengali, documents in Bengali).
We first apply ``bag of words'' exact term matching to retrieve a candidate set of documents.
Each document is split into sentences, and we apply inference with mBERT on each sentence separately.
The relevance score of each document is determined by combining the top $k$ scoring sentences with the document term-matching score as follows:
\begin{equation}
S_{doc} = \alpha \cdot S_{r} + 	(1-\alpha) \cdot \sum_{i=1}^{k} w_i \cdot S_i
\end{equation}
\noindent where $S_{i}$ is the $i$-th top sentence score according to BERT.
The parameters $\alpha$ and $w_i$'s can be tuned via cross-validation.
All candidate documents are resorted by the above score $S_{doc}$, which serves as the final output.

Our approach is a straightforward adaptation of the evidence combination technique of~\citet{Yilmaz_etal_EMNLP2019}, except using mBERT.
To be precise, we apply an mBERT model that has been fine-tuned on English relevance data {\it directly in the target language}, without any modification.

For the cross-lingual retrieval case, where, for example, the queries are in English and the documents are in French, we simply translate the query into the target language using Google Translate, and apply exactly the same methods as above.

\section{Experimental Setup}

\begin{table}
	\centering\resizebox{\columnwidth}{!}{
	\begin{tabular}{llll}
		\toprule
		Doc (Query) Language & Source & \# Topics & \# Docs \\
		\midrule
		Chinese~(zh, en)& NTCIR 8 & 73 & 308,832 \\
		Arabic~(ar, en) & TREC 2002 & 50 &383,872 \\
		French~(fr, en, zh) & CLEF 2006 & 49 & 171,109 \\
		Hindi~(hi) & FIRE 2012 & 50 & 331,599 \\
		Bengali~(bn) & FIRE 2012 & 50 & 500,122 \\
		English~(en, hi, bn) & FIRE 2012 & 50 &  392,577 \\
		\bottomrule
	\end{tabular}}
	\caption{Dataset Statistics.}
	\label{table:dataset}
\end{table}

\begin{table*}[t!]
	\small
	\centering\resizebox{\textwidth}{!}{
		\begin{tabular}{lll lll lll lll}
			\toprule
			
			& {\bf AP} & {\bf P@20} & {\bf NDCG@20}   & {\bf AP}  & {\bf P@20} & {\bf NDCG@20} & {\bf AP} & {\bf P@20} & {\bf NDCG@20}  \\
			\cmidrule(lr){2-4}  \cmidrule(lr){5-7}  \cmidrule(lr){8-10}
			{\bf Model} & \multicolumn{3}{c}{\textbf{NTCIR8-zh}} & \multicolumn{3}{c}{\textbf{TREC2002-ar}} & \multicolumn{3}{c}{\textbf{CLEF2006-fr}} \\
			
			\toprule
			BM25 & 0.4065 & 0.3911 & 0.4867 & 0.2923 & 0.3660 &0.4057 & 0.3111 & 0.3184 & 0.4458 \\
			\cmidrule{2-10}
			1S: $ \textrm{BERT}(\textrm{MB}) $ & 0.4466  & 0.4370 & 0.5288 & 0.3103 & 0.3940 & 0.4511 &  0.3115& 0.3255 & 0.4404  \\
			2S: $ \textrm{BERT}(\textrm{MB}) $ & 0.4587 & 0.4610 & 0.5577 & 0.3087 & 0.4000 & 0.4498 & 0.3347 & 0.3367 & 0.4639  \\
			3S: $ \textrm{BERT}(\textrm{MB}) $ & \textbf{0.4612} & \textbf{0.4651} & \textbf{0.5626} & \textbf{0.3105} & \textbf{0.4070} & \textbf{0.4547} & \textbf{0.3390} & \textbf{0.3429} & \textbf{0.4727} \\
			\cmidrule{2-10}
			tune-embed & 0.4458 & 0.4521 & 0.5443 & 0.3040 & 0.3860 & 0.4370 & 0.3064 & 0.3224 & 0.4396 \\
			\midrule 
			& \multicolumn{3}{c}{\textbf{FIRE2012-hi}} & \multicolumn{3}{c}{\textbf{FIRE2012-bn}} & \multicolumn{3}{c}{\textbf{FIRE2012-en}} \\
			\toprule
			BM25 & 0.3867 & 0.4470 & 0.5310 & 0.2881 & 0.3740 & 0.4261 & 0.3713 & 0.4970 & 0.5420 \\
			\cmidrule{2-10}
			1S: $ \textrm{BERT}(\textrm{MB}) $ & \textbf{0.4284} & \textbf{0.4750} & \textbf{0.5597} & 0.3210& 0.4130 & 0.4747 & 0.4424 & \textbf{0.5610} & 0.5971 \\
			2S: $ \textrm{BERT}(\textrm{MB}) $ & 0.4279 & 0.4740 & 0.5608 & 0.3228 & 0.4160 & 0.4802 & \textbf{0.4456} & \textbf{0.5610} & \textbf{0.6053} \\
			3S: $ \textrm{BERT}(\textrm{MB}) $ & 0.4259 & \textbf{0.4750} & 0.5590 & \textbf{0.3217} & \textbf{0.4190} & \textbf{0.4808} & 0.4432 & 0.5530 & 0.6008 \\
			\cmidrule{2-10}
			tune-embed & 0.4168 & 0.4720 & 0.5578 & 0.3086 & 0.4010 & 0.4606 & 0.4347 & 0.5400 & 0.5874 \\
			\bottomrule
		\end{tabular}
	}
	\caption{Mono-lingual ranking effectiveness.}
	\label{table:multilingual}
\end{table*}

\begin{table*}[t!]
	\small
	\centering\resizebox{\textwidth}{!}{
		\begin{tabular}{lll lll lll lll}
			\toprule
			
			& {\bf AP} & {\bf P@20} & {\bf NDCG@20}   & {\bf AP}  & {\bf P@20} & {\bf NDCG@20} & {\bf AP} & {\bf P@20} & {\bf NDCG@20}  \\
			\cmidrule(lr){2-4}  \cmidrule(lr){5-7}  \cmidrule(lr){8-10}
			{\bf Model} & \multicolumn{3}{c}{\textbf{NTCIR8-en-zh}} & \multicolumn{3}{c}{\textbf{TREC2002-en-ar}} & \multicolumn{3}{c}{\textbf{CLEF2006-en-fr}} \\
			
			\toprule
			BM25 & 0.2946 & 0.3260 & 0.3825 & 0.2678 & 0.3620 &0.3981 & 0.3070 & 0.3163 & 0.4476 \\
			\cmidrule{2-10}
			1S: $ \textrm{BERT}(\textrm{MB}) $ & 0.3289  & 0.3630 & 0.4233 & 0.2780 & 0.3620 & 0.4101 &  0.3152& 0.3306 & 0.4489  \\
			2S: $ \textrm{BERT}(\textrm{MB}) $ & 0.3416 & 0.3829 & 0.4443 & 0.2819 & 0.3590 & 0.4097 & 0.3349 & \textbf{0.3449} & 0.4783  \\
			3S: $ \textrm{BERT}(\textrm{MB}) $ & \textbf{0.3459} & \textbf{0.3945} & \textbf{0.4568} & \textbf{0.2853} & \textbf{0.3670} & \textbf{0.4175} & \textbf{0.3363} & 0.3439 & \textbf{0.4799} \\
			% \cmidrule{2-10}
			% CL-matching & 0.2994 & 0.3363 & 0.3908 & 0.2699 & 0.362 & 0.3991 & 0.3256 & 0.3316 & 0.4681 \\
			\midrule 
			& \multicolumn{3}{c}{\textbf{CLEF2006-zh-fr}} & \multicolumn{3}{c}{\textbf{FIRE2012-hi-en}} & \multicolumn{3}{c}{\textbf{FIRE2012-bn-en}} \\
			\toprule
			BM25 & 0.2274 & 0.2406 & 0.3428 & 0.3410 & 0.4600 & 0.4931 & 0.3044 & 0.4280 & 0.4637 \\
			\cmidrule{2-10}
			1S: $ \textrm{BERT}(\textrm{MB}) $ & 0.2351 & 0.2437 & 0.3470 & 0.3749& 0.4655 & 0.5014 & 0.3210 & 0.4163 & 0.4523 \\
			2S: $ \textrm{BERT}(\textrm{MB}) $ & 0.2524 & 0.2542 & 0.3703 & 0.3788 & 0.4750 & 0.5118 & \textbf{0.3308} & \textbf{0.4430} & \textbf{0.4836} \\
			3S: $ \textrm{BERT}(\textrm{MB}) $ & \textbf{0.2600} & \textbf{0.2656} & \textbf{0.3878} & \textbf{0.3817} & \textbf{0.4810} & \textbf{0.5188} & 0.3274 & 0.4395 & 0.4779 \\
			% \cmidrule{2-10}
			% CL-matching & 0.2316 & 0.2417 & 0.3506 & 0.3347 & 0.4489 & 0.4777 &0.2767 & 0.3935 & 0.4315 \\
			\bottomrule
		\end{tabular}
	}
	\caption{Cross-lingual ranking effectiveness.}
	\label{table:cross}
\end{table*}

As previously discussed, we examined two different retrieval tasks:\ mono-lingual retrieval in the target language and cross-lingual retrieval.
Dataset statistics are summarized in Table~\ref{table:dataset}.
For each corpus, we indicate the query language(s); the queries are in parallel if multiple languages are provided.
All these languages are captured in mBERT and are from diverse language families (Sino-Tibetan, Semitic, Romance, and Indo-Aryan).

For mono-lingual retrieval, we examined the following conditions:\
NTCIR 8 IR4QA Track~(Simplified Chinese), %\footnote{\url{http://research.nii.ac.jp/ntcir/permission/ntcir-8/perm-en-ACLIA.html}}
TREC 2002 CLIR Track~(Arabic), %\footnote{\url{https://trec.nist.gov/pubs/trec11/t11\_proceedings.html}}
CLEF 2006 Ad-Hoc Track~(French), %\footnote{\url{http://clef.isti.cnr.it/2006/2006Ad-hoc.html}} and
FIRE 2012 Ad-Hoc Track~(Bengali, Hindi, English). %\footnote{\url{https://www.isical.ac.in/~fire/2012/index.html}}
In each case, the document and query languages are the same, indicated in parentheses.

For cross-lingual retrieval, we examined the following conditions:\
NTCIR 8 IR4QA Track~(English $\rightarrow$ Simplified Chinese),
TREC 2002 CLIR Track~(English $\rightarrow$ Arabic),
CLEF 2006 Ad-Hoc Track~(\{English, Chinese\} $\rightarrow$ French),  and
FIRE 2012 Ad-Hoc Track~(\{Bengali, Hindi\} $\rightarrow$ English).
In all cases above, the notation ($X \rightarrow Y$) indicates that the queries are in language $X$ and that the documents are in language $Y$.

For model fine-tuning, we followed basically the same experimental setup as~\citet{Yilmaz_etal_EMNLP2019}.
We used data from the Microblog Tracks from TREC 2011--2014~\cite{Lin_etal_TREC2014}, setting aside 75\% of the total data for training and the rest for validation, which is used for selecting the best model parameters. 
We trained the model using cross-entropy loss with a batch size of 16; the Adam optimizer is applied with an initial learning rate of $ 3 \times 10^{-5}$. 
During fine-tuning, the embeddings are not updated for better cross-lingual generalization ability, which we empirically show.

For inference (e.g., document ranking), Google Translate is first used to translate queries into the language of the documents (in the case of cross-lingual retrieval).
The query is then used to retrieve the top 1000 hits from the corpus using BM25 as the ranking function.
For this, we used the open-source Anserini IR toolkit~\cite{Yang_etal_JDIQ2018}\footnote{\url{http://anserini.io/}} with minor modifications based on version 0.6.0 to swap in Lucene Analyzers for different languages.
Fortunately, Lucene provides analyzers for all the languages in our test collections.
In all cases, we used the default BM25 parameters in Anserini.

We use average precision (AP), precision at rank 20 (P@20), and NDCG@20 as the evaluation metrics.
Following \citet{Yilmaz_etal_EMNLP2019}, we considered up to the top three sentences in aggregating sentence-level evidence.
We also applied five-fold cross-validation on all datasets and the parameters $\alpha$ and the $w_i$'s were obtained by grid search, choosing the parameters that yield the highest AP.

\section{Results and Discussion}

Our results are shown in Table~\ref{table:multilingual} (mono-lingual) and Table~\ref{table:cross} (cross-lingual).
The top row of each section shows the effectiveness of the BM25 baseline.
The remaining blocks show the effectiveness of our models;
the $n$S preceding the model name indicates that inference was performed using the top $n$ scoring sentences from each document.

From Table~\ref{table:multilingual}, we find that mBERT fine-tuned on the microblog data outperforms the BM25 baseline by a large margin for all three metrics, for all collections.
It is worth emphasizing that the model was not fine-tuned with any of the corpora used in retrieval.
These results indicate that mBERT effectively transfers its relevance matching ability across languages, from English to Chinese, Arabic, French, Hindi, and Bengali.
Furthermore, note that the test collections are all from the news domain, while the training data are drawn from social media.
This implies that mBERT is able to transfer relevance matching models across domains and across languages simultaneously.

Note that one of the FIRE2012 conditions is English, which provides a sanity check for these experiments; here, we reproduce the gains observed by~\citet{Yilmaz_etal_EMNLP2019}.
Also consistent with previous work, looking at the $n$S configurations, we see that using {\it only} the top-scoring sentence already yields a high level of effectiveness, showing that the best sentence alone provides a good proxy of document relevance.
Adding the second or third sentence yields small improvements at best.

We see different degrees of effectiveness gains across languages:\ for some languages (e.g., Chinese), we observe a large gain; for others (e.g., Arabic and French), the gains are more modest.
Beyond making this observation, we currently have no explanation why.
These differences might arise from intrinsic language differences in mBERT (i.e., the pretraining regime), characteristics of the test collection (e.g., types of queries), differences in the Anserini document processing pipeline (e.g., tokenization), or likely, a combination of all these factors (and more).
We save an in-depth exploration of this question for future work.

To support our modeling decision to fix the token embeddings of mBERT during fine-tuning, we experimented with a contrastive condition in which the embeddings were fine-tuned as well.
This is shown in the entry ``tune-embed'' in Table~\ref{table:multilingual}.
Although we conducted experiments using the three different sentence configurations, only the best results are shown for space considerations.
Comparing these results with the fixed-embedding setting, we observe that fine-tuning the embeddings leads to lower effectiveness.
We suspect that allowing the embeddings to change alters the underlying cross-lingual relationship between tokens from different languages, because the English token embeddings are updated while those from other languages remain unchanged.
As a result, it is possible that mBERT learns a relevance matching model that is more specific to English, affecting its ability to transfer to other languages.

For the cross-lingual setting, results from Table~\ref{table:cross} are consistent with the mono-lingual results.
Recall that the only difference here is our use of Google Translate to translate the query into the document language.
Note that the BM25 baselines are lower than in the mono-lingual case, especially for the NTCIR8-en-zh, CLEF2006-zh-fr, and TREC2002-en-ar conditions, with drops of 0.1119, 0.0837, and 0.0245 in AP, respectively.

Error analysis attributes the issue to the use of Google Translate as an imperfect black box translator. 
For example, we have the NTCIR query ``\textit{Who is Lung Yingtai?}''\ (a famous writer and poet).
The correct Chinese translation is \begin{CJK}{UTF8}{gbsn}``谁是\underline{龙应台}？''\end{CJK} but with Google Translate, we obtain \begin{CJK}{UTF8}{gbsn} ``\underline{隆应泰}是谁？''\end{CJK}\ (a totally different person).
Such translation errors are expected because of the lack of context and background knowledge. 
On the other hand, the CLEF2006-en-fr condition has a much smaller effectiveness drop for the BM25 baseline because the English and French queries share some tokens, such as person names.

However, these results show that, even with imperfect top one translations, we observe substantial gains in cross-lingual and cross-domain relevance transfer.
This suggests that better ways of query translation, for example, taking advantage of multiple translations~\cite{Ture_Lin_TOIS2104}, represents a promising approach.

\section{Conclusion}

Building on two recent papers~\cite{Yilmaz_etal_EMNLP2019,wu-dredze-2019-beto}, we empirically show that mBERT is able to transfer models of relevance matching cross-linguistically, {\it without any special processing}.
This is empirically supported by document retrieval experiments in five different languages drawn from diverse language families.
For the mono-lingual (non-English) case, we can rerank documents retrieved using ``bag of words'' exact term matching directly with mBERT.
For the cross-lingual case, we find that Google Translates provides an adequate, albeit imperfect, black box solution to translate the query language into the document language.

Our findings open up lots of interesting questions regarding language differences, which will drive future work.
However, we believe our most impactful contribution is highlighting a potential avenue for building high-quality search engines for low(er)-resources languages by leveraging relevance judgments in languages where they are far more plentiful.

\section*{Acknowledgments}

This research was supported by the Natural Sciences and Engineering Research Council (NSERC) of Canada, and enabled by computational resources provided by Compute Ontario and Compute Canada.

%\bibliography{clir}
\bibliographystyle{acl_natbib}

\end{document}